# Butterfly Patterns for Stretched Inhomogeneous Gel Networks using Large-Scale Molecular Dynamics Simulations

Katsumi Hagita[a]* and Takahiro Murashima[b]

[a]Department of Applied Physics, National Defense Academy, 1-10-20 Hashirimizu, Yokosuka 239-8686, Japan
[b]Department of Physics, Tohoku University, 6-3 Aramaki-aza-Aoba, Aoba-ku, Sendai 980-8578, Japan
*Corresponding Author Email: hagita@nda.ac.jp

**ABSTRACT:** Large-scale coarse-grained molecular dynamics simulations of inhomogeneous gel networks were performed to investigate abnormal butterfly patterns in two-dimensional scattering. The networks were diamond lattice-based with distributions in the number of beads between the crosslink points. Remarkably, the results confirm that the abnormal butterfly pattern originates from stronger inhomogeneity. For the examined systems, the range of scattering wavevector $q$ for the normal butterfly pattern was markedly different from those for the abnormal butterfly patterns. The findings address an essential aspect of the discrepancy between theorical prediction and experimental observations.

Butterfly patterns have been observed in scattering experiments involving various polymer materials and have received considerable attention in the investigation of nanoscale structural changes. Nearly 30 years ago, many studies [1–11] on "butterfly patterns" attempted to reveal the origin of abnormal butterfly patterns that represent enhanced scattering in the stretching direction. Macroscopic theory predicts normal butterfly patterns with peaks perpendicular to the stretching direction. Hashimoto [2] explained the dark streak in the direction perpendicular to stretching in the small-angle region as a significant behavior of the (abnormal) butterfly pattern. They investigated polymers under flow using small-angle light-scattering experiments and Fourier transform processing of the observed real-space images. Onuki [4] proposed a theoretical model to predict the crossover from normal butterfly patterns (theoretical prediction) to abnormal butterfly patterns (experimental observation) using the irregularity parameter of the crosslinked structure. However, the theoretical prediction has not been conclusively verified because microscopic verification is infeasible for real gel materials.

Our recent work [12] developed a bond-breakable model of coarse-grained molecular dynamics (CGMD) simulations based on the Kremer‑Grest model [13], and a model of an inhomogeneous gel network by introducing bimodal distributions of the bead number between crosslink points based on the diamond lattice. This bimodal distribution causes the inhomogeneity in the gel network. Consequently, we show that the shape of the stress–strain (SS) curve near the yield point becomes convex, which is closer to experimental observations. We used a phantom chain model that allows chain crossing for comparison with our previous work [14]. We confirmed that the two-dimensional scattering patterns (2DSPs) of the stretched gel network showed normal butterfly patterns and bright streaks that originate from total failure, although we were unable to obtain detailed information regarding the 2DSPs in the small-angle region owing to their small system size and poor resolution.

In this study, we performed large-scale CGMD simulations using the Kremer–Grest model, which prohibits chain crossing. The simulations reveal that the abnormal butterfly pattern in the small-angle region depends on the inhomogeneity of the network. We show that, in the networks examined, normal and abnormal butterfly patterns occur in different ranges of scattering wavevector $q$. The findings contribute to understanding the discrepancy between theorical prediction and experimental observations of abnormal butterfly patterns, which has not been addressed in previous studies.

As explained in our previous work [12], we used LAMMPS [15,16] to perform the stretching CGMD simulations. In the evaluation of 2DSPs, uniaxial anisotropy was assumed, as well as various polymer material models [12, 17–22]. A system with a diamond tetragonal lattice and 16 × 16 × 16 unit cells was placed under periodic boundary conditions at $\rho = 0.1$. We used $N_s = 25$ as a uniform network, where $N_s$ indicates the number of particles between the bonded crosslink points. A bimodal network is defined as a network wherein two lengths ($N_s \pm N_b$) are randomly assigned with equal probability as the number of particles between cross-linking points. We investigated two bimodal net-works with $N_b$ values of 10, 12, and 15. The total numbers of particles and bonds are listed in Table I.

Table. 1 Total number of particles and bonds

| $N_b$ | Number of particles | Number of bonds |
|---|---|---|
| 0 | 1,671,168 | 1,703,936 |
| 10 | 1,673,948 | 1,706,716 |
| 12 | 1,674,782 | 1,707,550 |
| 15 | 1,675,338 | 1,708,106 |

In the stretching CGMD simulations, we applied a stretching rate of $d\varepsilon/dt = 10^{-4}/\tau$ in the $x$-direction and the bond-breakable model [12] with $r_{bb} = 1.47$. Our previous work [12], we used the quartic potential originally used by Stevens [23,24] and developed the parameters for the quartic potential.

Figure 1 shows stress–strain (SS) curves for $N_b$ = 0, 10, 12, and 15. For a large $N_b$ (inhomogeneity), the shape of the stress–strain (SS) curve becomes convex near the yield point, which is consistent with our previous work [12]. Figure 2 shows the 2DSPs just before the yield point. For $N_b$ = 0, only a normal butterfly pattern was observed. For $N_b$ = 15, we observed an enhancement in scattering (abnormal butterfly pattern) in the stretching direction in the small-angle region. These results show that the $q$ range of the abnormal butterfly pattern was significantly different from that of the normal butterfly pattern.

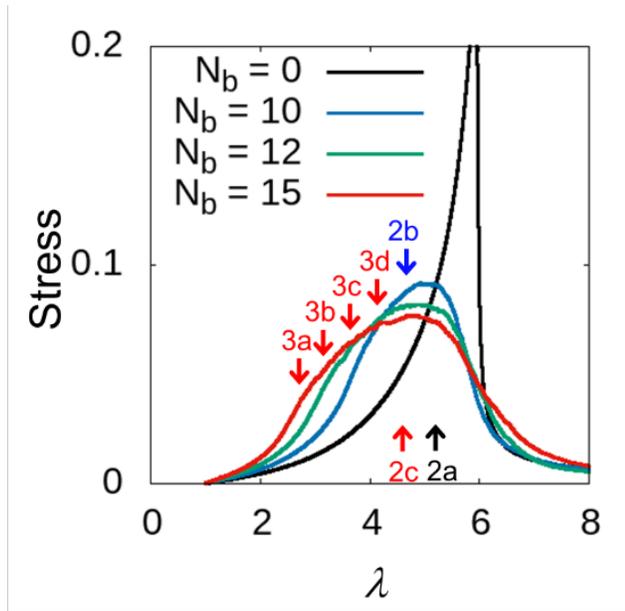

Figure 1. Stress–strain curves for $N_b$ = 0, 10, 12, and 15. The arrows indicate the strains of the 2DSPs shown in Figures 2 and 3.

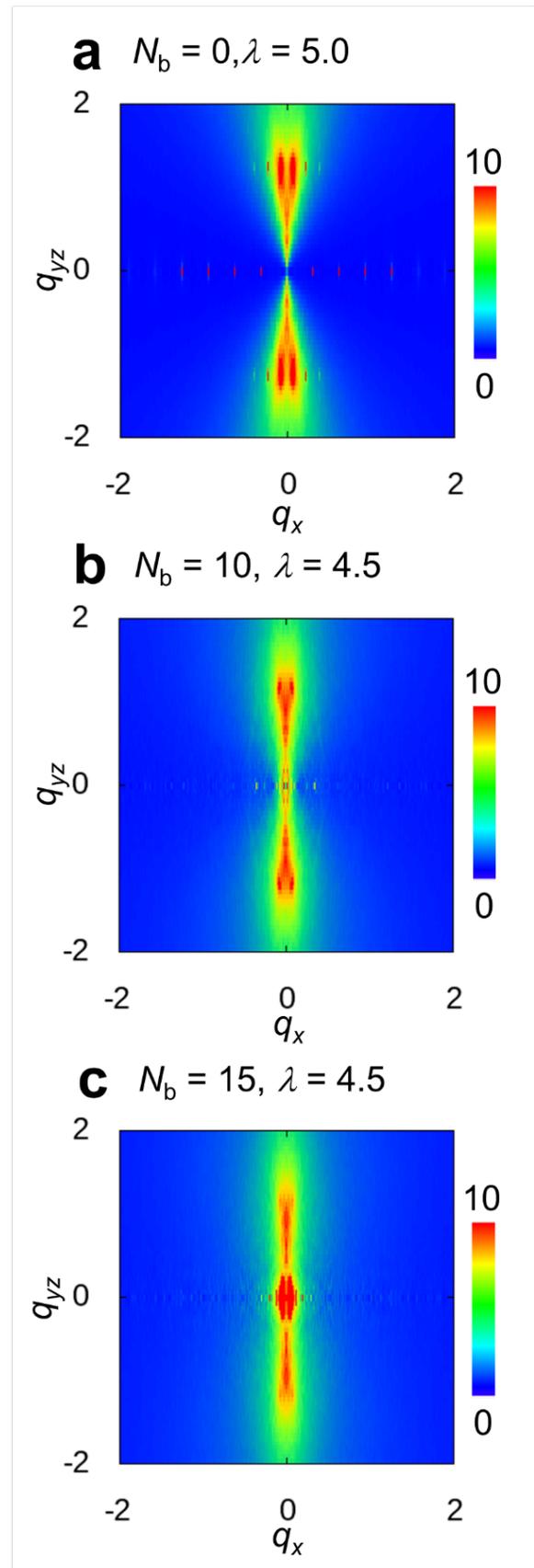

Figure 2. 2DSPs just before the yield point for $N_b$ = 0, 10, and 15.

Figure 3 shows the 2DSPs for $N_b$ = 15. For $\lambda$ = 2.5, no abnormal butterfly pattern was observed. From $\lambda$ = 3.0 to 4.0, the abnormal butterfly patterns are observed to increase in intensity, but the peak positions remain almost the same. The magnified 2DSPs clearly show dark streaks in the direction perpendicular to stretching in the small-angle region.

[4]. To track detailed changes of the peak $q$ positions and intensity distributions against inhomogeneity $N_b$, the 2DSP-resolution should be improved, i.e., larger simulations with a larger system size should be conducted in future work.

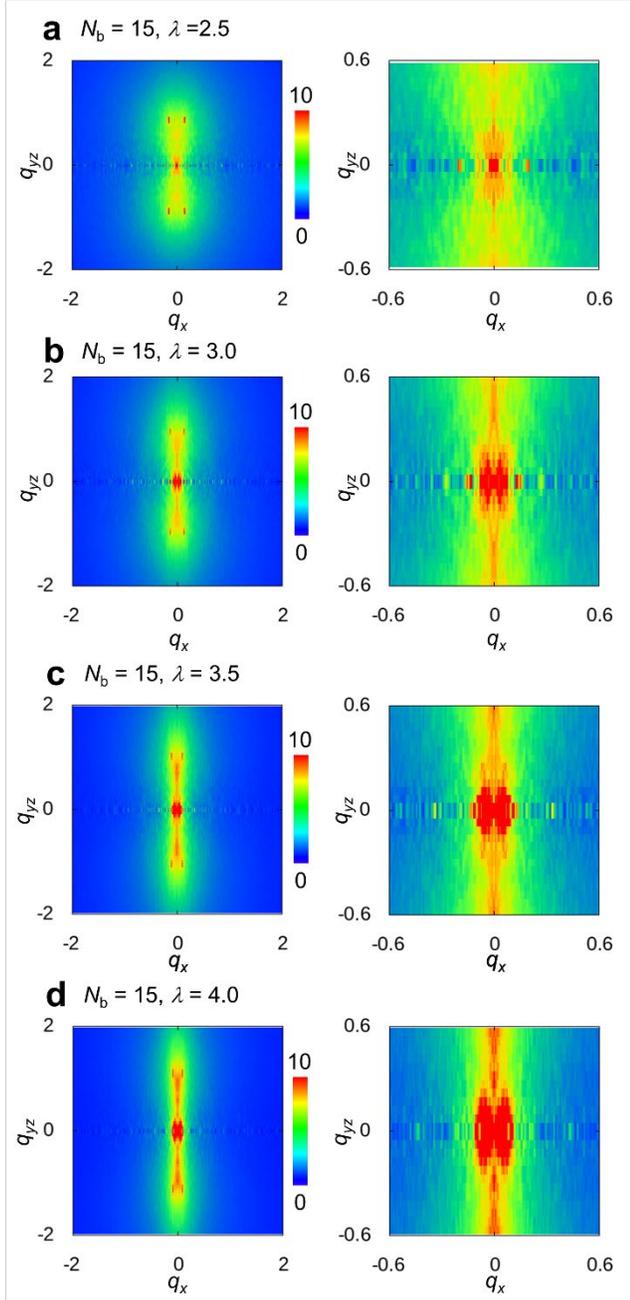

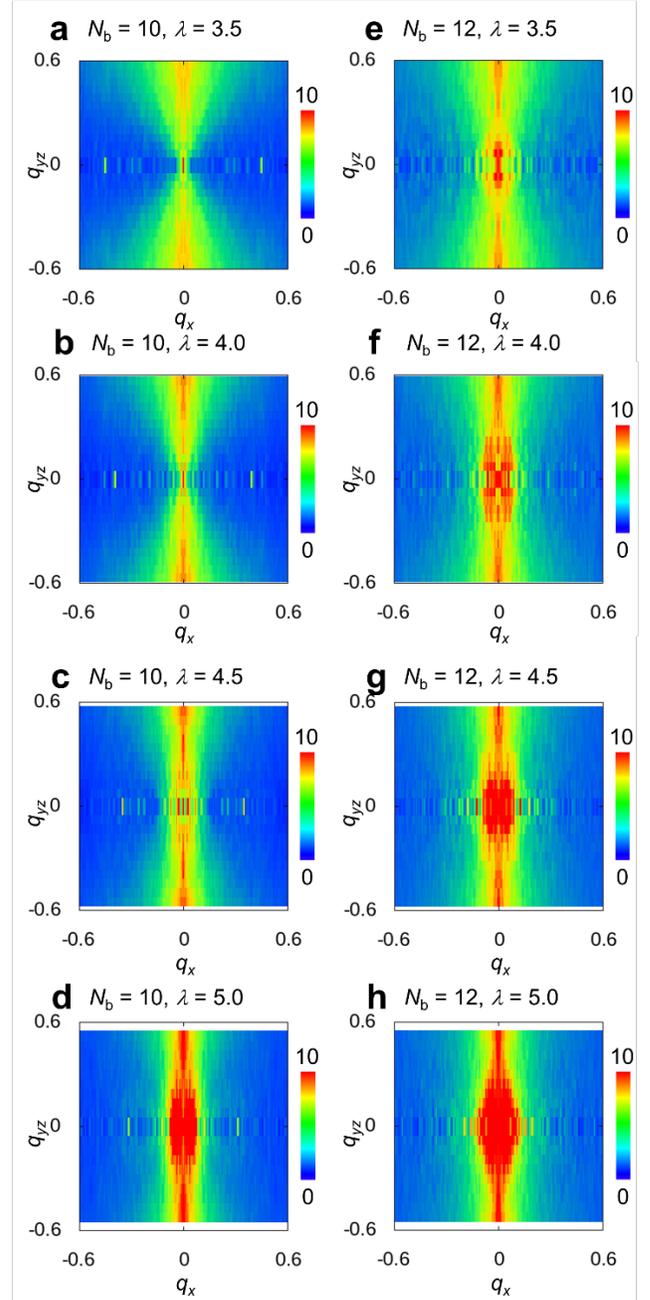

Figure 3. 2DSPs at $\lambda$ = 2.5 to 4.0 for $N_b$ = 15. Magnified 2DSPs with the same color range are plotted on the right.

Figure 4. Magnified 2DSPs at $\lambda$ = 3.5 to 5.0 for $N_b$ = 10 and 12.

Figure 4 shows the 2DSPs for $N_b$ = 10 and 12. Abnormal butterfly patterns were observed for $\lambda$ = 4.5 and 5.0. The $q_x$ values of the peaks of the abnormal butterfly patterns increased with $N_b$. This is consistent with Onuki's theory

In conclusion, we confirmed that an inhomogeneous network showing abnormal butterfly patterns can be reproduced via large-scale CGMD simulations using the Kremer–Grest model with bond breakable potentials. Here, we introduced the inhomogeneity using distributions of the number of beads between crosslink points in a diamond lattice-based network. We confirmed that the stronger the inhomogeneity, the stronger the abnormal butterfly pattern. In the investigated system, the $q$ range of the normal butterfly pattern was significantly different from that of the abnormal butterfly patterns. This informs the understanding of the discrepancy between theorical prediction and experimental observations for abnormal butterfly patterns.

## AUTHOR INFORMATION

### Corresponding Author

*Email: hagita@nda.ac.jp

### Author Contributions

The manuscript was written with contributions from all authors. The conceptualizations, computations, analyses, and visualizations were mainly performed by K. H. All authors approved the final version of the manuscript.

### Notes

The authors declare no competing financial interest.

## ACKNOWLEDGMENT

The authors thank Dr. T. Indei for their valuable discussions. This study used the computational resources of the High-Performance Computing Infrastructure (HPCI) in Japan, project nos. hp220215, hp230148, hp230384, and hp240052. The authors were partially supported by JSPS KAKENHI, Japan (grant numbers JP18H04494, JP20H04649, and JP21H00111) and JST CREST, Japan (grant number JPMJCR1993).

SYNOPSIS TOC

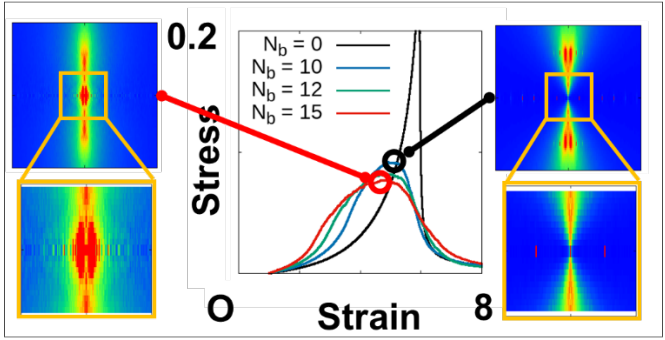